\documentclass{ifacconf}
\usepackage{amsmath,amssymb} 
\usepackage{natbib}
\usepackage{graphicx}      
\usepackage{xcolor}
\usepackage{float}
\usepackage{fancyhdr}

\usepackage{fancyhdr}
\DeclareMathOperator*{\argmin}{arg\,min}
\DeclareMathOperator*{\trace}{tr}
\fancypagestyle{firstpage}
{
    \fancyhead[L]{This work has been submitted to IFAC for possible publication.}    
    \fancyhead[R]{}
}
\begin{document}
\begin{frontmatter}

\title{Control Contraction Metric Synthesis for Discrete-time Nonlinear Systems}  

\author{Lai Wei,} 
\author{Ryan Mccloy,} 
\author{Jie Bao\thanksref{footnoteinfo}}
\thanks[footnoteinfo]{Corresponding author: Jie Bao. This work was supported by ARC Discovery Grant DP180101717.}

\address{School of Chemical Engineering, The University of New South Wales, NSW 2052, Australia (lai.wei1@unsw.edu.au, r.mccloy@unsw.edu.au, j.bao@unsw.edu.au)}

\begin{abstract}
Flexible manufacturing has been the trend in the area of the modern chemical process nowadays. One of the essential characteristics of flexible manufacturing is to track time-varying target trajectories (e.g. diversity and quantity of products). A possible tool to achieve time-varying targets is contraction theory. However, the contraction theory was developed for continuous time systems and there lacks analysis and synthesis tools for discrete-time systems. This article develops a systematic approach to discrete-time contraction  analysis  and  control  synthesis  using  Discrete-time Control Contraction Metrics (DCCM) which can be implemented using Sum of Square (SOS) programming. The proposed approach is demonstrated by illustrative example. 
\end{abstract}


\end{frontmatter}
\section{Introduction}
\thispagestyle{firstpage}

Traditionally a chemical plant is designed for and operated at a certain steady-state operating condition, where the plant economy is optimised. Nowadays, the supply chains are increasingly dynamic. Process industry needs to shift from the traditional mass production to more agile, cost-effective and dynamic process operation, to produce products of different specifications to meet the market demand, and deal with the variations in specifications, costs and quantity of supplied raw material and energy. As such, operational flexibility has become one key feature of the next-generation “smart plants”, i.e., the control systems need to be able to drive the process systems to any feasible time-varying operational targets (setpoints)  in response to the dynamic supply chains.

While most chemical processes are nonlinear, linear controllers have been designed based on linearised models. This approach is defensible for regulatory control around a predetermined steady state. Flexible process operation warrants nonlinear control as the target operating conditions can vary significantly to optimise the economic cost. Existing nonlinear control methods, e.g., Lyapunov-based approaches typically require redesigning the control Lyapunov function and controller when the target  equilibrium changes, not suitable for flexible process operation with time varying targets. This has motivated increased interest for alternative approaches based on the contraction theory framework \citep{WangBao17, wang2017distributed}. Introduced by \cite{lohmiller1998contraction}, contraction theory facilitates stability analysis and control of nonlinear systems with respect to arbitrary, time-varying (feasible) references without redesigning the control algorithm \citep{manchester2017control,lopez2019contraction}. Instead of using the state space  process model alone, contraction theory also exploits the differential dynamics, a concept borrowed from fluid dynamics, to analyse the local stability of systems. Thus, one useful feature of contraction theory is that it can be used to analyse the incremental stability/contraction of nonlinear systems, and synthesise a controller that ensures offset free tracking of feasible target trajectories using control contraction metrics (or CCMs, see, e.g., \cite{manchester2017control}).  As most process control systems are developed and implemented in a discrete time setting \citep{goodwin2001control}, warranting the tools for analysing, designing and implementing contraction-based control for discrete-time systems. However, the current contraction-based control synthesis (e.g., \cite{manchester2014control,manchester2017control,manchester2018robust}), is limited to continuous-time control-affine nonlinear systems.

The contributions of this work include the developments of discrete-time control contraction metrics (DCCMs) and a systematic analysis, control synthesis and implementation of contraction-based control for discrete-time nonlinear systems. The structure of this article is as follows. In Section~\ref{review}, we introduce the concept of contraction theory in the context of discrete-time nonlinear systems and the concept of sum of squares (SOS) programming. Section~\ref{DCCM} presents the main results - the systematic approach to discrete-time contraction analysis and control synthesis using DCCMs. Section~\ref{SIM} illustrates the details of numerical computation and implementation of the proposed contraction-based approach using an example of CSTR control, followed by the conclusions.  

\textbf{Notation.} Denote by $f_k = f(x_k)$ for any function $f$, $\mathbb{Z}$ represents the set of all integers, $\mathbb{Z}^+$ represents set of positive integers, $\mathbb{R}$ represents set of real numbers. $\Sigma$ represents a polynomial sum of squares function that is always non-negative, e.g. $\Sigma(x_k,u_k)$ is a polynomial sum of squares function of $x_k$ and $u_k$.

\section{Background}\label{review}
\subsection{Overview of Contraction Theory}
\begin{figure}[tb]
    \begin{center}
    \includegraphics[width=8.4cm]{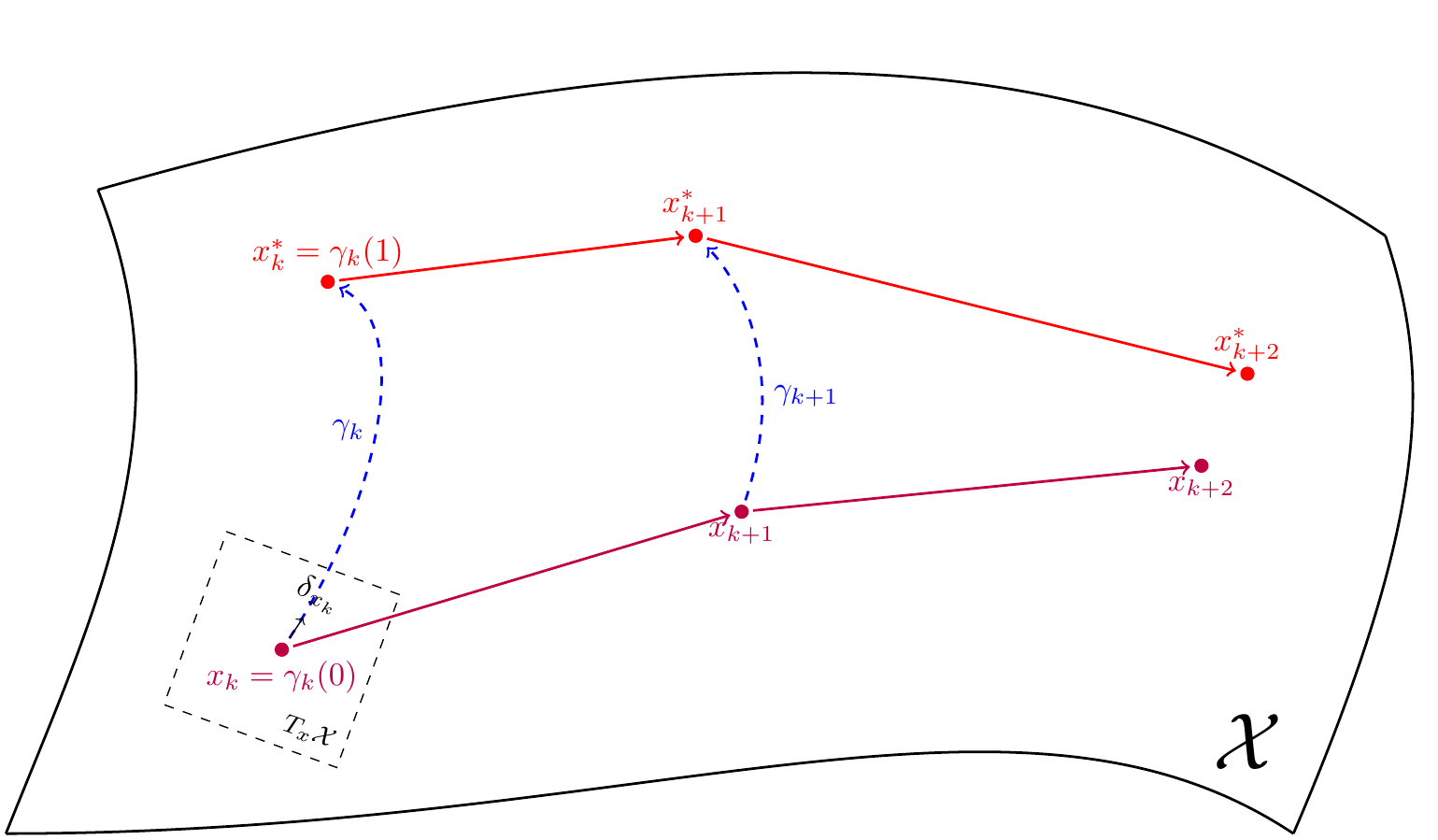}
    \caption{Illustration for contracting of discrete-time systems.} 
    \label{fig:geodesic}
    \end{center}
 \end{figure}

The contraction theory \citep{lohmiller1998contraction,manchester2017control} provides a set of analysis tools for the convergence of trajectories with respect to each other via the concept of displacement dynamics or \textit{differential dynamics}. Fundamentally, under the contraction theory framework, the evolution of distance between any two infinitesimally close neighbouring trajectories is used to study the distance between any finitely apart pair of trajectories. Consider a discrete-time nonlinear system of the form 
\begin{equation}\label{equ:dynamic equation}
    x_{k+1} = f(x_k),
\end{equation}
where $x_k \in \mathcal{X} \subseteq \mathbb{R}^n$ is the state vector at time step $k\in \mathbb{Z}^+$. 

Consider two neighbouring discrete-time trajectories separated by an infinitesimal displacement $\delta_{x_k}$. Formally, $\delta_{x_k} := \frac{\partial x_k}{\partial s}$ is a vector in the tangent space $T_x\mathcal{X}$ at $x_k$, where $s$ parameterises the state trajectories of~\eqref{equ:dynamic equation} (see, e.g., Figure~\ref{fig:geodesic}). The discrete-time differential dynamics of system \eqref{equ:dynamic equation} are then defined as
\begin{equation}\label{equ:differential dynamical system}
	\delta_{x_{k+1}}= \frac{\partial f(x_k)}{\partial x_k}\delta_{x_k},
\end{equation}where $\frac{\partial f(x_k)}{\partial x_k}$ is the Jacobian matrix of $f$ at $x_k$. 

From Riemannian geometry \citep{do1992riemannian}, a state-dependent matrix function, $\Theta$, can be used to define a generalised infinitesimal displacement, $\delta_{z_k}$, where
\begin{equation}\label{equ:coordinate transformation}
    \delta_{z_k} = \Theta(x_k)\delta_{x_k}.
\end{equation} 
The infinitesimal squared distance for system (\ref{equ:differential dynamical system}) is described by $\delta_{x_k}^T\delta_{x_k}$. Furthermore, a generalisation of infinitesimal squared distance, $V_k$, is defined using the coordinate transformation \eqref{equ:coordinate transformation}, i.e.
\begin{equation}\label{equ:generalised distance}
    V_k= \delta_{z_k}^T\delta_{z_k} = \delta_{x_k}^T\Theta(x_k)^T\Theta(x_k)\delta_{x_k} := \delta_{x_k}^TM_k\delta_{x_k},
\end{equation}where the symmetric positive definite matrix function $M_k :=\Theta(x_k)^T\Theta(x_k)$ is uniformly bounded, i.e.
\begin{equation}
    \alpha_1 I \leq M(x_k) \leq \alpha_2 I
\end{equation}
for some constants $\alpha_2 \geq \alpha_1 > 0$. 

\begin{defn}[\cite{lohmiller1998contraction}]\label{def:contraction region}
    System \eqref{equ:dynamic equation} is in contraction region with $0 < \beta \leq 1$, with respect to a uniformly bounded metric function, $M$, provided
    \begin{equation}\label{inequ:contraction condition}
        \begin{aligned}
            \delta^T_{x_{k}} (A_k^T M_{k+1} A_k - M_{k})\delta_{x_{k}}&\leq -\beta \delta_{x_k}^TM_{k}\delta_{x_k} < 0,
        \end{aligned}
    \end{equation}
    for $\forall x, \delta_x \in \mathbb{R}^n$. Furthermore, the infinitesimal squared distance, $V_k$ in \eqref{equ:generalised distance}, can be viewed as a discrete-time differential Lyapunov function, i.e.
    \begin{equation}\label{eqe:Vdotcondition}
    V_{k+1} - V_k \leq - \beta V_k < 0, \quad \forall x, \delta_x \in \mathbb{R}^n.
    \end{equation}
\end{defn}

\begin{thm}[\cite{lohmiller1998contraction}]\label{thm:original}
    Given system \eqref{equ:control affine}, any trajectory which starts in a ball of constant radius with respect to the metric $M$, centred at a given trajectory and contained at all times in a generalised contraction region, remains in that ball and converges exponentially to this trajectory.
\end{thm}
The above theorem will be used subsequently to develop a discrete-time contraction-based control approach. 

Since contraction theory is built on top of Riemannian geometry, thus several concepts of Riemannian geometry need to be clarified. For a smooth curve, $c: s \in [0,1] \to \mathcal{X}$, connecting the two points, $x_0, x_1 \in \mathcal{X}$ (i.e. with $c(0) = x_0$, $c(1) = x_1$), we define the corresponding Riemannian distance, $d(x_0,x_1)$, and energy, $E(x_0,x_1)$, as
\begin{equation}
\begin{split}
\label{equ:Rdist_Renergy}
d(x_0,x_1) :=\! \int_0^1 \sqrt{\delta_{c(s)}^\top M(c(s))) \delta_{c(s)})} \, ds, \\ E(x_0,x_1) :=\! \int_0^1 \delta_{c(s)}^\top M(c(s)) \delta_{c(s)} \, ds.
\end{split}
\end{equation}
Furthermore, the minimum length curve or geodesic, $\gamma$, connecting any two points, e.g. $x_0,x_1 \in \mathcal{X}$, is defined as
\begin{equation}\label{equ:geodesic}
\gamma(x_0,x_1) := \argmin_c d(x_0,x_1).
\end{equation}
\begin{rem}\label{rmk:decreasing length}
According to \cite{lohmiller1998contraction}, system \eqref{equ:dynamic equation} is a contracting system if a metric function $M$ exists such that condition \eqref{inequ:contraction condition} holds (in this case $M$ is a \textit{contraction metric}). Morevover, the Riemannian distance and energy, as $d(x_0,x_1)$ and $E(x_0,x_1)$ in  \eqref{equ:Rdist_Renergy} respectively, for any two points $(x_0,x_1)$ can both be shown to decrease exponentially under the evolution of the system \eqref{equ:dynamic equation} (see also \citep{manchester2017control}).  
\end{rem}

Control contraction metrics or \textit{CCMs} generalise the above contraction analysis for autonomous systems to the controlled system setting, whereby the analysis searches jointly for a controller and the metric that describes the contraction properties of the resulting closed-loop system. Herein, we will consider 
discrete-time control-affine nonlinear systems of the form
\begin{equation}\label{equ:control affine}
    x_{k+1} = f(x_k) + g(x_k)u_k,
\end{equation}
where $x_k \in \mathcal{X} \subseteq  \mathbb{R}^n$ is the state, $u_k \in \mathcal{U} \subseteq  \mathbb{R}^m$ is the control input, and $f$ and $g$ are smooth functions.
The corresponding differential dynamics of (\ref{equ:control affine}) are defined as
\begin{equation}\label{equ:differential control affine}
    \delta_{x_{k+1}} = A(x_k)\delta_{x_k} + B(x_k)\delta_{u_k},
\end{equation}
where $A:=\frac{\partial (f(x_k) + g(x_k)u_k)}{\partial x_k}$ and $B:=\frac{\partial (f(x_k) + g(x_k)u_k)}{\partial u_k}$ are Jacobian matrices, and $\delta_{u_k} := \frac{\partial u_k}{\partial s}$ is a vector in the tangent space $T_x\mathcal{U}$ at $u_k$.

\cite{manchester2017control} showed that the existence of a CCM for a continuous-time nonlinear system was sufficient for globally stabilising every forward-complete solution of that system. It can be extended to discrete-time systems, i.e., for discrete-time control contraction metrics (DCCMs) consider the discrete-time equivalent differential state feedback control law as in \citep{manchester2017control}
\begin{equation}\label{equ:differential feedback}
    \delta_{u_k} = K(x_k) \delta_{x_k},
\end{equation}
and integrate~\eqref{equ:differential feedback} along the geodesic, $\gamma$~\eqref{equ:geodesic}, i.e., one particular feasible tracking controller, can be defined as
\begin{equation}\label{equ:control integral}
    u_k = u^*_k + \int_0^1K(\gamma(s))\frac{\partial \gamma(s)}{\partial s}ds.
\end{equation}
where the target trajectory sequence $(x_k^*,u_k^*,x_{k+1}^*)$ is a solution sequence for the nonlinear system \eqref{equ:control affine}, and $\gamma$ is a geodesic joining $x$ to $x^*$ (see~\eqref{equ:geodesic}). Note that this particular formulation is target trajectory independent, since the target trajectory variations do not require structural redesign. Moreover, the discrete-time control input, $u_k$~\eqref{equ:control integral}, is a function with arguments, $(x_k,x^*_k,u^*_k)$, and hence the current state and target trajectory give the current input.  

In summary, a suitably designed contraction-based controller ensures that the length of the minimum path (i.e., geodesic) between any two trajectories (e.g., the plant state, $x_k$, and desired state, $x^*_k$, trajectories), with respect to the metric $M_k$, shrinks with time. We next introduce sum of squares programming to provide one possible means for computing DCCMs and the corresponding feedback gains, as required for contraction-based control.

\subsection{Sum of Squares (SOS) Programming}
Sum of squares programming (see, e.g. \cite{boyd2004convex}) was proposed as a tractable method in \citep{manchester2017control} for computing CCMs for continuous-time control-affine nonlinear systems and will be demonstrated in this article as an additionally tractable approach in the discrete-time setting. A polynomial $p(x)$, is a SOS polynomial, provided it satisfies
\begin{equation}\label{equ:SOS definition}
    p(x) = \sum_{i=1}^n{q_i(x)^2},
\end{equation} where $q_i(x)$ is a polynomial of $x$. Thus, it is easy to see that any SOS polynomial, $p$, is positive provided it can be expressed as in \eqref{equ:SOS definition}. Furthermore, in \citep{aylward2006algorithmic}, determining the SOS property expressed in \eqref{equ:SOS definition} is equivalent to finding a positive semi-definite $Q$ such that 
\begin{equation}
    p(x)=v(x)^TQv(x) \in \Sigma (x),
\end{equation}where $v(x)$ is a vector of monomials that is less or equal to half of the degree of polynomial $p(x)$, $\Sigma$ is stated in Notation. An SOS programming problem is an optimisation problem to find the decision variable $p(x)$ such that $p(x)\in\Sigma (x)$. In~\cite{Parrilo00}, the non-negativity of a polynomial is determined by solving  a semi-definite programming (SDP) problem (e.g. SeDuMi~\citep{Sturm1999}).


\subsection{Problem Summary and General Approach}
The contraction condition for autonomous discrete-time systems is described in \eqref{inequ:contraction condition}, which requires additional considerations for control, especially with respect to contraction-based controller synthesis. In the following section, this condition will be explicitly stated for control-affine systems \eqref{equ:control affine}, by incorporating the differential controller \eqref{equ:differential feedback} into the contraction condition \eqref{inequ:contraction condition}. Inspired by continuous-time SOS synthesis approaches, which search jointly for a contraction metric and controller, subsequent sections detail the transformation of the contraction condition for a discrete-time control-affine nonlinear system into a tractable SDP problem, solvable by SOS programming. 




\section{Contraction Metric and Controller Synthesis for Discrete-time Nonlinear Systems}\label{DCCM}
This section presents the transformation of a contraction condition for discrete-time control-affine nonlinear systems into a tractable synthesis problem, solvable via SOS programming. 
\subsection{Obtaining a Tractable Contraction Condition}
Substituting the differential dynamics (\ref{equ:differential control affine}) and differential feedback control law (\ref{equ:differential feedback}) into the contraction condition (\ref{inequ:contraction condition}), we have the immediate discrete-time  result.
\begin{lem}\label{lemma}
    A discrete-time control-affine nonlinear system \eqref{equ:control affine} is contracting (i.e. stabilisable to feasible reference trajectories) if a pair of functions $(M,K)$, where $M$ \eqref{equ:generalised distance} is a positive definite DCCM and $K$ is a differential feedback \eqref{equ:differential feedback}, exists such that following condition holds for all $x$
    \begin{equation}\label{inequ: control affine contraction condition}
        (A_k+B_kK_k)^TM_{k+1}(A_k+B_kK_k) - (1-\beta)M_{k} < 0.
    \end{equation}
\end{lem}

As characterised by Lemma \ref{lemma}, two conditions are needed to ensure contraction of discrete-time nonlinear systems -- the first is the discrete-time contraction condition \eqref{inequ: control affine contraction condition}, and the second is the positive definite property of the metric $M$. Inspired by \citep{manchester2017control}, an equivalent condition to \eqref{inequ: control affine contraction condition} is developed in the following Theorem as a tractable means for handling the bilinear terms in \eqref{inequ: control affine contraction condition}. 

\begin{thm}\label{thm:condition}
    Consider a differential feedback controller \eqref{equ:differential feedback} for the differential dynamics \eqref{equ:differential control affine} of a discrete-time control-affine nonlinear system \eqref{equ:control affine} and denote the inverse of the metric matrix as $W:=M^{-1}$. Then, the discrete-time nonlinear system \eqref{equ:control affine} is contracting with respect to a DCCM, $M$, if a pair of matrix functions $(W,L)$ satisfies 
    \begin{equation}\label{inequ:LMI}
        \begin{bmatrix}
            W_{k+1}              && A_kW_k+B_kL_k \\
            (A_kW_k+B_kL_k)^T    && (1-\beta)W_k
        \end{bmatrix} > 0,
    \end{equation}
    where $A_k,B_k$ and $K_k$ are functions in \eqref{equ:control affine} and \eqref{equ:differential feedback} respectively, $W_{k+1}:= M_{k+1}^{-1}= M^{-1}(f(x_k)+B(x_k)u_k)$, $L_k := K_kW_k$ and $\beta \in (0,1]$.
\end{thm}
\begin{pf}
Condition \eqref{inequ: control affine contraction condition} is equivalent to 
\begin{equation}\label{Metric_Condi}
    (1-\beta)M_{k} - (A_k+B_kK_k)^TM_{k+1}(A_k+B_kK_k) > 0.
\end{equation}
Applying Schur's complement \citep{boyd1994linear} to (\ref{Metric_Condi}) yields
\begin{equation}
\label{equ:schursCC}
    \begin{bmatrix}
        M_{k+1}^{-1} && (A_k+B_kK_k) \\
        (A_k+B_kK_k)^T     && (1-\beta)M_k
    \end{bmatrix} > 0.
\end{equation}
Defining $W_k := M^{-1}(x_k)$ and $W_{k+1} := M^{-1}(x_{k+1}) = M^{-1}(f(x_k)+B(x_k)u_k)$, we then have
\begin{equation}
\label{equ:inequality_step}
    \begin{bmatrix}
        W_{k+1}     && (A_k+B_kK_k) \\
        (A_k+B_kK_k)^T    && (1-\beta)W_k^{-1}
    \end{bmatrix} > 0.
\end{equation}
Left/right multiplying \eqref{equ:inequality_step} by an invertible positive definite matrix, yields
\begin{equation}
    \begin{bmatrix}
        I   && 0 \\
        0   && W_k
    \end{bmatrix}^T
    \begin{bmatrix}
        W_{k+1}     && (A_k+B_kK_k) \\
        (A_k+B_kK_k)^T    && (1-\beta)W_k^{-1}
    \end{bmatrix}
    \begin{bmatrix}
        I   && 0 \\
        0   && W_k
    \end{bmatrix}  > 0,
\end{equation}
which is equivalent to the following condition
\begin{equation}
    \begin{bmatrix}
        W_{k+1}          && (A_k+B_kK_k)W_k \\
        W_k^T(A_k+B_kK_k)^T    && (1-\beta)W_k
    \end{bmatrix} > 0.
\end{equation}
Finally, defining $L_k:=K_kW_k$, we have the condition \eqref{inequ:LMI}. 
\end{pf}
\begin{rem}
\label{rem:LMI}
The rationale behind transforming the contraction condition in Lemma \ref{lemma} into Theorem \ref{thm:condition} lies in obtaining the contraction metric $M$ and feedback controller gain $K$. Since in equation \eqref{inequ: control affine contraction condition} there are terms coupled with several unknowns, e.g. $B_kK_k^TM_{k+1}B_kK_k$, their computation becomes an incredibly difficult (if not intractable) problem. Applying Schur's complement, whilst decoupling these terms, introduced an inverse term, namely $M_{k+1}^{-1}$, and an additional coupled term in \eqref{equ:schursCC}, which were handled via reparameterisation. Consequently, this motivates the development of the equivalent contraction condition in \eqref{inequ:LMI} that effectively removes these computational complexities.
\end{rem}
Suppose $(M,L)$ are found satisfying \eqref{inequ:LMI}, then, the corresponding differential feedback controller in \eqref{equ:differential feedback} can be reconstructed using $K_k = L_kW_k^{-1}$ and hence, the controller in \eqref{equ:controller} can be computed.

Naturally, the next step is to demonstrate that the pair $(M,L)$ satisfying \eqref{inequ:LMI} can be computed. The following section will demonstrate how the contraction condition of Theorem \ref{thm:condition} is computationally tractable, through the use of one particular approach -- SOS programming, i.e., how the inequality in equation \eqref{inequ:LMI} can be used to obtain the DCCM $M$ and corresponding feedback gain $K$ that are required for implementing a contraction-based controller.

\subsection{Synthesis via SOS Programming}
In this section, we explore an SOS programming method, as one possible approach to obtaining a DCCM and feedback control law, which satisfy the contraction requirements outlined in the previous sections. First, we present SOS programming with relaxations for Lemma \ref{lemma}, followed by some discussion and natural progression to SOS programming for Theorem \ref{thm:condition}.

From Lemma \ref{lemma}, two conditions need to be satisfied: the contraction condition \eqref{inequ: control affine contraction condition} and positive definite property of the matrix function $M$. These conditions can be transformed into an SOS programming problem (see, e.g.,~
\cite{Parrilo00,aylward2006algorithmic}) if we assume the functions are all polynomial functions or polynomial approximations (see, e.g.,~\cite{EbeAll06}), i.e. 
\begin{equation}
    \begin{aligned}
        & \min_{k_c,m_c} \trace(M)\\
        & s.t. ~ v^T \Omega_1 v  \in \Sigma (x_k,u_k,v), v^T M_k v \in \Sigma (x_k,v),
    \end{aligned}
\end{equation}where $\Omega_1 = -((A_k+B_kK_k)^TM_{k+1}(A_k+B_kK_k) - (1-\beta)M_{k})$ represents the discrete-time contraction condition and $k_c,m_c$ are coefficients of polynomials for the controller gain, $K$ in \eqref{equ:differential feedback}, and metric, $M$, respectively (see the example in Section \ref{SIM} for additional details). This programming problem is computationally difficult (if not intractable) due to the hard constraints imposed by the inequality \eqref{inequ: control affine contraction condition}. One possible improvement can be made by introducing relaxation parameters to soften the constraints (see, e.g. \citep{boyd2004convex}), i.e. introducing two small positive values, $r_1$ and $r_2$, as
\begin{equation}
    \begin{aligned}
        & \min_{k_c,w_c,r_1,r_2} r_1 + r_2 \\
        & s.t. ~ v^T \Omega_1 v - r_1I \in \Sigma (x_k,u_k,v)\\
        & v^T M_k v -r_2I \in \Sigma (x_k,v),r_1 \geq 0,r_2 \geq 0.
    \end{aligned}
\end{equation}
Note that the contraction condition holds if the two relaxation parameters $r_1$ and $r_2$ are some positive value, then we get a required DCCM as long as the relaxation parameters are positive. Although this relaxation reduces the programming problem difficulty, the problem remains infeasible, due to the terms coupled with unknowns, e.g. $B_kK_k^TM_{k+1}B_kK_k$. 


Naturally, substitution for the equivalent contraction condition \eqref{inequ:LMI} in Theorem \eqref{thm:condition}, solves this computational obstacle (see Remark \ref{rem:LMI}), and hence a tractable SOS programming problem can be formed as follows
\begin{equation}\label{min:sos}
    \begin{aligned}
        &\underset{l_c,w_c,r}{\min} r\\
        &s.t. \ w^T \Omega w -rI \in \Sigma(x_k,u_k,w), ~r \geq 0,
    \end{aligned}
\end{equation} 
where $\Omega =         
\begin{bmatrix}
    W_{k+1}              && A_kW_k+B_kL_k \\
    (A_kW_k+B_kL_k)^T    && (1-\beta)W_k
\end{bmatrix}$ and $l_c$ represents the polynomial coefficients of $L_k$ (see the example in Section \ref{SIM} for additional details). Note that the inverse and coupling terms are not present in \eqref{min:sos} and that its SDP tractable solution yields the matrix function $W_k$ and $L_k$, as required for contraction analysis and control.

Finally, the controller \eqref{equ:control integral} can be constructed, following online computation of the geodesic \eqref{equ:geodesic} using the metric $M_k = W_k^{-1}$ and feedback gain $K_k = L_kW_k^{-1}$.


In conclusion, by considering the differential controller \eqref{equ:differential feedback}, the contraction condition \eqref{inequ:contraction condition} for control of \eqref{equ:control affine} was make explicit in Lemma \ref{lemma}. Then, a tractable equivalent condition was framed by Theorem \ref{thm:condition} which was shown to be transformed into the SOS programming problem \eqref{min:sos}. Consequently, the contraction metric $M$ and stabilising feedback gain $K$ can be obtained by solving \eqref{min:sos} using an SDP tool such as SeDuMi, resulting in a tractable synthesis approach for contraction-based control of the discrete-time nonlinear system \eqref{equ:control affine}.

\section{Illustrative Example}\label{SIM}
Section \ref{DCCM} presented Theorem \ref{thm:condition} as a computationally feasible contraction condition, that can be transformed into the tractable SOS programming problem \eqref{min:sos}. In this section, we will show how to calculate the geodesic numerically and present a case study to illustrate the synthesis and implementation of a contraction-based controller. 
\subsection{Numerical Geodesic and Controller Computation}
Suppose the optimisation problem \eqref{min:sos} is solved for the pair $(W,L)$ and hence $(M,K)$ are obtained. The next step in implementing the contraction-based controller \eqref{equ:control integral} is to integrate \eqref{equ:controller} along the geodesic, $\gamma$ \eqref{equ:geodesic}. 
Subsequently, one method to numerically approximate the geodesic is shown. From \eqref{equ:Rdist_Renergy} and \eqref{equ:geodesic} we have the following expression for computing the geodesic, 
\begin{equation}\label{GEO_CAL}
    \gamma(x_0,x_1) = \argmin_c \int_0^1{\frac{\partial c(s)}{\partial s}^T M(c(s)) \frac{\partial c(s)}{\partial s}ds}.
\end{equation}

Since  \eqref{GEO_CAL} is  an  infinite  dimensional  problem  over  all smooth curves, without explicit analytical solution, the problem must be discretised to be numerically solved. Note that the integral can be approximated by discrete summation provided the discrete steps are sufficiently small. As a result, the geodesic (\ref{GEO_CAL}) can be numerically calculated via the following optimisation problem
\begin{equation}\label{min:geodesic}
    \begin{aligned}
    \bar \gamma(x_0,x_1) = \argmin_c &\sum_{i=1}^N{\Delta x_{s_i}^T M(x_i) \Delta x_{s_i} \Delta s_i}&\\
    s.t.  &\sum_{i=1}^N \Delta x_{s_i} \Delta s_i = x_1 - x_0&\\
          &\sum_{i=1}^N \Delta s_i = 1,\quad  x_i = \sum_{j=1}^i \Delta x_{s_j} \Delta s_j&
    \end{aligned}
\end{equation}where $\bar \gamma(x_0,x_1) \approx \gamma(x_0,x_1)$ represents the numerically approximated geodesic, $x_0$ and $x_1$ are the endpoints of the geodesic, $\Delta x_{s_i} := \Delta_{x_i} / \Delta_{s_i}$ where $\Delta x_{s_i}$ can be interpreted as the displacement vector discretised with respect to the $s$ parameter, $\Delta {s_i}$ is a small positive scalar value and $N$ is the chosen discretisation step size, where $x_i$ represents the numerical state evaluation along the geodesic. 
\begin{rem}
Compare \eqref{min:geodesic} with \eqref{GEO_CAL} and note that $\Delta x_{s_i}$ and $\Delta_{s_i}$ represent the discretisations of  $\frac{\partial c(s)}{\partial s}$ and $\delta_s$ respectively. Furthermore, note that the second and third constraints in \eqref{min:geodesic} ensure that the integral from $s=0$ to $s=1$ aligns with the discretised path connecting the start, $x_0$, and end, $x_1$, state values. Hence, as $\Delta_{s_i}$ approaches 0, i.e. for an infinitesimally small discretisation step size, the approximated discrete summation in \eqref{min:geodesic} converges to the smooth integral in \eqref{GEO_CAL}.
\end{rem}
After the geodesic is numerically calculated using \eqref{min:geodesic}, the control law in \eqref{equ:control integral} can be analogously calculated using an equivalent discretisation as follows
\begin{equation}\label{equ:controller}
    u_k = u_k^* +  \sum_{i=1}^N \Delta x_{s_i} \Delta s_i K(x_i).
\end{equation}
\begin{rem}
    Observe that the values required to implement this control law are calculated from optimisation \ref{min:geodesic}. For example, recall that $x_i$ represents the numerical state evaluation along the geodesic at discrete points. Substituting $K_k = L_kW_k^{-1}$ (see optimisation \ref{min:sos}) into \eqref{equ:controller}, we can then implement the control law as follows
\end{rem}
\begin{equation}
\label{equ:disccontr}
    u_k = u_k^* +  \sum_{i=1}^N  \Delta x_{s_i} L_kW_k^{-1} \Delta s.
\end{equation}

\subsection{Control of a CSTR}
In this section, we will demonstrate the synthesis and implementation of a discrete-time contraction-based controller via an illustrative example. Consider the following unitless discrete-time nonlinear system model for a CSTR (adapted from \cite{KUBICEK1980987})
\begin{equation}
\label{equ:CSTR}
    x_{k+1} = 
    \begin{bmatrix}
        \begin{aligned}
            &1.1x_{1_k} - 0.1 x_{1_k}x_{2_k} + u_k\\
            &0.9x_{2_k} + 0.1 x_{1_k}
        \end{aligned}
    \end{bmatrix}.
\end{equation}
To obtain the corresponding differential system model (see \eqref{equ:differential control affine}), the Jacobian matrices are calculated as follows:
\begin{equation}
    A_k = 
    \begin{bmatrix}
        1.1-0.1x_{2_k} & -0.1x_{1_k}\\
        0.1 & 0.9
    \end{bmatrix},
    B_k = 
    \begin{bmatrix}
        1\\
        0
    \end{bmatrix}.
\end{equation}

As required for \eqref{min:sos}, we predefine the respective contraction metric and feedback gain duals, $W_k$ and $L_k$, as matrices of polynomial functions, i.e.
\begin{equation}
\label{equ:WkLk}
    W_k = 
    \begin{bmatrix}
        W_{{11}_k} && W_{{12}_k} \\
        W_{{12}_k} && W_{{22}_k}
    \end{bmatrix}, \qquad
    L_k =
    \begin{bmatrix}
        L_{1_k} \\
        L_{2_k}
    \end{bmatrix},
\end{equation}where $W_{{\cdot \cdot}_k}=w_{{\cdot \cdot c}}v(x_k)$ and $w_{{\cdot \cdot c}}$ is a row vector of unknown coefficients, and similarly for $L_k$. As required to solve the SOS problem in \eqref{min:sos}, the functions $W_{{\cdot \cdot k}}$ etc. need to be polynomial functions and as such are expressed using the common monomial vector, $v(x_k)$, defined as 
\begin{equation}
    v(x_k) = 
    \begin{bmatrix}
        x_{1_k}^6 & x_{1_k}^5x_{2_k} & \cdots & x_{2_k}^6
    \end{bmatrix}^T,
\end{equation}
where the polynomial order is chosen to be 6. 
Additionally, $W_{k+1}$, is defined as a matrix of polynomials
\begin{equation}
    W_{k+1} = 
    \begin{bmatrix}
        W_{{11}_{k+1}} && W_{{12}_{k+1}} \\
        W_{{12}_{k+1}} && W_{{22}_{k+1}}
    \end{bmatrix},
\end{equation}where the elements are defined as $W_{{\cdot \cdot}_{k+1}}=w_{{\cdot\cdot c}}v(x_{k+1})$ and $w_{{\cdot \cdot c}}$ is the same coefficient vector for $W_k$ in \eqref{equ:WkLk}. 

Choosing $\beta=0.1$, the corresponding convergence rate  with respect to the DCCM, $M_k$ is equal to $0.9$ (see \eqref{eqe:Vdotcondition}). The SOS programming problem in \eqref{min:sos} can then be solved using MATLAB with YALMIP \citep{Lofberg2004} and SeDuMi \citep{Sturm1999} for the coefficient vectors $W_k$ and $L_k$ \eqref{equ:WkLk} (numerical values are provided in Appendix \ref{app:coeff}). 

At any time instant, $k$, we can compute the approximated geodesic, $\Delta x_{s_i}$, between $x_k$ and $x_k^*$ via the minimisation problem \eqref{min:geodesic}, using $M_k = W_k^{-1}$ and choosing a constant $\Delta_s = 1/N = 1/30$ (i.e., choosing to partition  Riemannian paths into 30 discrete segments). A contraction-based controller \eqref{equ:disccontr} can then be implemented.

\begin{figure}
  \begin{center}
  \includegraphics[width=\linewidth]{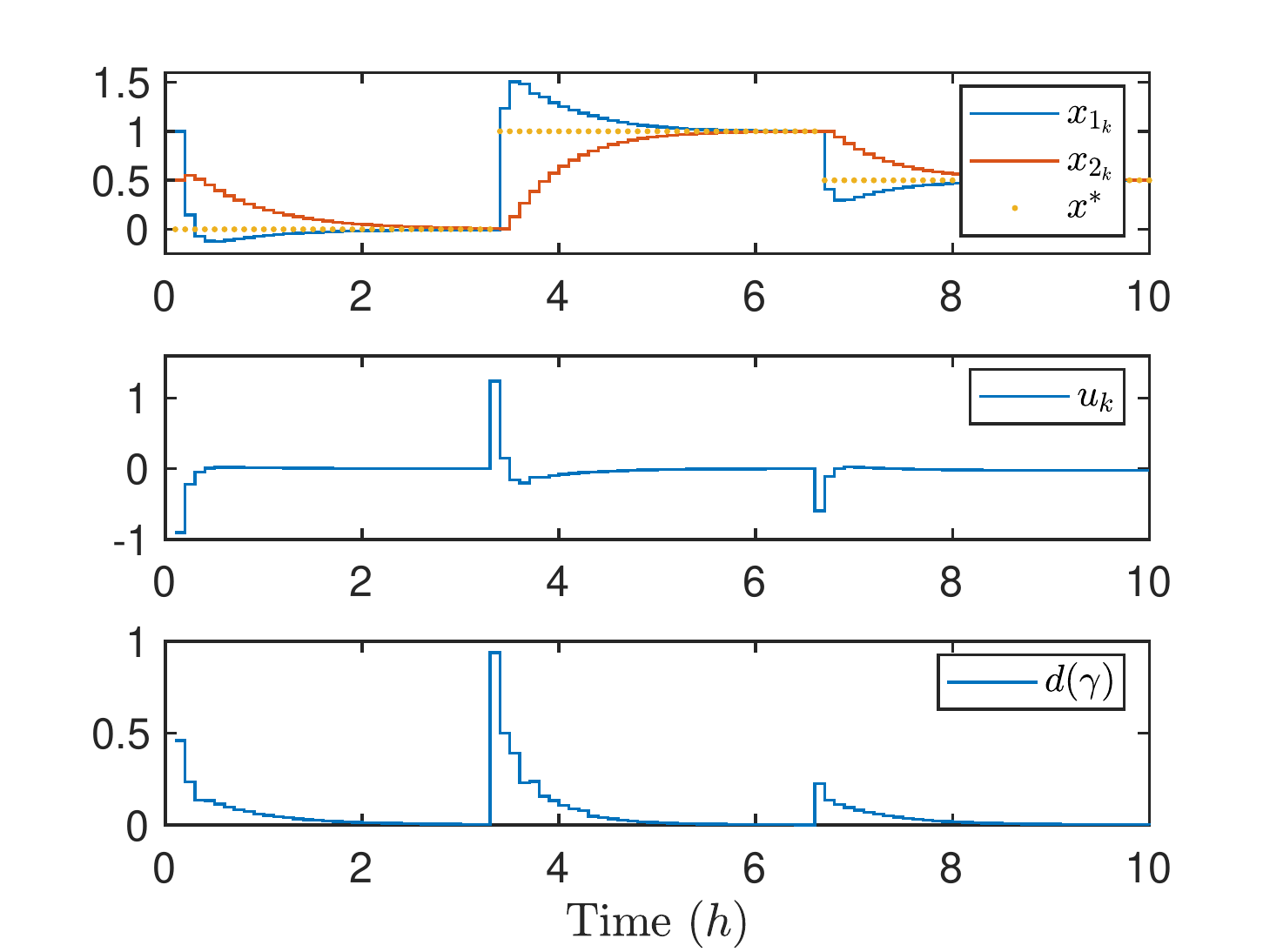}  
  \caption{System \eqref{equ:CSTR} -- state reference, $x^*$, and response, $x$, (top), contraction-based control input, $u$, \eqref{equ:disccontr} (middle) and geodesic length, $d(\gamma)$, \eqref{equ:Rdist_Renergy}, \eqref{min:geodesic} (bottom).}
  \label{fig:state}
  \end{center}
\end{figure}

To demonstrate the reference-independent tracking capabilities of a contraction-based controller, the CSTR \eqref{equ:CSTR} was simulated to track the time-varying reference $x_1^*=x_2^*=x^* = 0,1,0.5$ on the respective intervals $k T_s = [0,3.3),[3.3,6.6),[6.6,10]$, with sampling period $T_s = 0.1$, using the discrete-time contraction-based controller \eqref{equ:disccontr}. The  corresponding control reference values, $u^*= 0, 0, -0.025$, are calculated by solving the system model \eqref{equ:CSTR} at steady state. The resulting state response (top) and control effort (middle) are shown in Figure \ref{fig:state}. Observe that the system state tracks the time-varying reference without error and without structural controller redesign. Furthermore, Figure \ref{fig:state} (bottom) illustrates that the geodesic length, $d(\gamma)$, (see \eqref{equ:Rdist_Renergy}, \eqref{min:geodesic}), decreases exponentially following instantaneous reference changes for the controlled system as expected (see Remark \ref{rmk:decreasing length}). 
\section{Conclusion}\label{conclusion}
A systematic approach to discrete-time contraction  analysis  and  control  synthesis  using  Discrete-time Control Contraction Metrics (DCCM) was developed in this paper. By considering the differential controller \eqref{equ:differential feedback}, the contraction condition \eqref{inequ:contraction condition} for discrete-time control-affine nonlinear systems \eqref{equ:control affine} was derived in Lemma \ref{lemma}. A computationally tractable equivalent condition was framed by Theorem \ref{thm:condition}, which was transformed into an SOS programming problem \eqref{min:sos}. Consequently, the contraction metric, $M$, and stabilising feedback gain, $K$, were obtainable by solving \eqref{min:sos} using an SDP tool, resulting in a synthesis approach for contraction-based control of discrete-time control-affine nonlinear systems. Numerical computation for the geodesic and controller was described. The proposed approach was illustrated by a case study of CSTR control.

\bibliography{ifacconf}             
\appendix

\section{Coefficients of Matrix Functions in Section~\ref{SIM}} 
\label{app:coeff}
The polynomial coefficient vectors of $W$ and $L$ used in the Illustrative Example Section \ref{SIM} are as follows

$w_{{11c}}$ =
   [4.5868
   -0.0237
    0.1742
    2.0684
    0.1005
    2.7412
   -0.0038
   -0.0333
    0.1304
    0.2714
    2.4268
   -0.2897
    2.1171
    0.0132
    6.9634
    0.0005
    0.0150
   -0.0203
    0.0001
    0.0031
    0.0000
    0.0001
    0.0006
    0.0934
    0.0116
    0.0234
   -0.0000
    0.0000],

$w_{{12c}}$ =
    [-1.8328
     0.0654
    -0.1007
    -0.3102
    -0.0460
    -2.5568
     0.0001
     0.0177
    -0.0427
    -0.0515
    -0.2710
     0.0327
    -0.2791
     0.0207
    -1.3630
    -0.0000
    -0.0015
     0.0048
    -0.0120
     0.0013
     0.0000
     0.0000
    -0.0001
    -0.0092
    -0.0054
    -0.0031
    -0.0000
     0.0000],

$w_{{22c}}$ =
    [7.2139
    -0.0124
     0.0012
     0.0618
     0.0954
     1.1859
     0.0000
    -0.0034
     0.0088
     0.0296
     0.0303
    -0.0002
     0.0377
     0.0987
     0.4190
     0.0000
     0.0001
    -0.0010
     0.0016
    -0.0007
    -0.0000
     0.0000
     0.0000
     0.0013
     0.0012
     0.0059
    -0.0000
     0.0000],
 
     $l_{1c}$ =
   [-3.3514
    -0.0118
     0.2920
    -2.0838
     0.1256
    -1.6818
     0.0136
    -0.2138
     0.1707
     0.0306
    -2.6709
     0.3296
    -2.2965
     0.1873
    -4.8506
    -0.0282
     0.2366
    -0.1405
     0.5670
    -0.1170
     0.6971
    -0.0001
    -0.0009
    -0.0998
    -0.0076
    -0.0427
     0.0003
     0.0000],

     $l_{2c}$ =
    [0.1711
     0.6323
    -0.1381
     0.2945
    -0.4221
     0.3728
     0.0011
     0.0007
     0.0482
    -0.3245
     0.2982
    -0.0407
     0.2632
    -0.3511
     0.4786
     0.0031
    -0.0261
     0.0159
    -0.0331
     0.0474
    -0.1364
     0.0000
     0.0001
     0.0097
     0.0053
     0.0016
     0.0001
     0.0000].


\end{document}